\documentclass[journal,comsoc]{IEEEtran}
\usepackage[T1]{fontenc}

\usepackage{subfig}
\usepackage{caption}
\usepackage{float}
\usepackage{setspace}
\usepackage{graphicx}
\usepackage{graphicx,dblfloatfix}
\usepackage{tikz}
\usepackage{color}
\usepackage{amsmath}
\usepackage{amssymb} 
\usepackage{amsthm}

\usepackage{maplestd2e}
\usetikzlibrary{shapes,arrows,shadows}
\usepackage{xcolor}
\usepackage{lipsum}

\ifCLASSINFOpdf
\else
\fi
\usepackage{amsmath}
\usepackage[cmintegrals]{newtxmath}
\begin{document}
%

\title{A Spectrally Efficient Linear Polarization Coding Scheme for Fiber Nonlinearity Compensation in CO-OFDM Systems}

%
%
%

\author{O.~S.~Sunish Kumar,~\IEEEmembership{Student Member,~IEEE,}
	O.~A.~Dobre,~\IEEEmembership{Fellow,~IEEE,}
	~R.~Venkatesan,~\IEEEmembership{Life Senior Member,~IEEE,}
	S.~K.~Wilson,~\IEEEmembership{Fellow,~IEEE,}
	O.~Omomukuyo,~\IEEEmembership{Member,~IEEE},
	A.~Amari,~\IEEEmembership{Member,~IEEE}, 
	D.~Chang,~\IEEEmembership{Member,~IEEE}
	\thanks{O. S. Sunish Kumar, O. A. Dobre, R. Venkatesan, O. Omomukuyo, A. Amari, and D. Chang are with the Department of Computer Engineering,  Memorial University, St. John’s, NL, A1B 3X5, Canada, e-mail: skos71@mun.ca.}
	\thanks{S. K. Wilson is with Electrical Engineering Department, Santa Clara University, 500 El Camino
		Real, Santa Clara, CA 95053, United States.}}
\maketitle

\begin{abstract}
In this paper, we propose a linear polarization coding scheme (LPC) combined with the phase conjugated twin signals (PCTS) technique, referred to as LPC-PCTS, for fiber nonlinearity mitigation in coherent optical orthogonal frequency division multiplexing (CO-OFDM) systems. The LPC linearly combines the data symbols on the adjacent subcarriers of the OFDM symbol, one at full amplitude and the other at half amplitude. The linearly coded data is then transmitted as phase conjugate pairs on the same subcarriers of the two OFDM symbols on the two orthogonal polarizations. The nonlinear distortions added to these subcarriers are essentially anti-correlated, since they carry phase conjugate pairs of data. At the receiver, the coherent superposition of the information symbols received on these pairs of subcarriers eventually leads to the cancellation of the nonlinear distortions. We conducted numerical simulation of a single channel 200 Gb/s CO-OFDM system employing the LPC-PCTS technique. The results show that a Q-factor improvement of 2.3 dB and 1.7 dB with and without the dispersion symmetry, respectively, when compared to the recently proposed phase conjugated subcarrier coding (PCSC) technique, at an average launch power of 3 dBm. In addition, our proposed LPCPCTS technique shows a significant performance improvement when compared to the 16-quadrature amplitude modulation (QAM) with phase conjugated twin waves (PCTW) scheme, at the same spectral efficiency, for an uncompensated transmission distance of 2800 km.
\end{abstract}


%
\IEEEpeerreviewmaketitle

\vspace{-0.2cm}
\section{Introduction}
%
%
%
%
\IEEEPARstart{T}{he} fiber Kerr nonlinearity imposes an upper limit on the maximum achievable transmission capacity in coherent optical orthogonal frequency division multiplexing (CO-OFDM) systems \cite{REssiambre}. There have been extensive efforts to surpass the Kerr nonlinearity limit through several mitigation techniques. Recently, a technique called phase conjugated twin wave (PCTW) has been proposed for effective mitigation of the nonlinear distortions in a polarization multiplexed optical transmission system \cite{XLiu}. In this technique, mutually phase conjugated twin waves are transmitted on the two orthogonal polarizations and are coherently superimposed at the receiver. In such a transmission scheme, the nonlinear distortions imposed onto two PCTWs, which are modulated onto the same optical carrier, are essentially anti-correlated, when a dispersion symmetry condition is met in the link. Note that the dispersion symmetry in the transmission link can be readily obtained by performing a 50\% electronic dispersion pre-compensation (pre-EDC) at the transmitter. At the receiver, the coherent superposition of the received PCTWs thus leads to the cancellation of the nonlinear distortions, albeit at the expense of halving the overall spectral efficiency of the link. In \cite{XYi}, a variant of PCTW technique for COOFDM system exploring the Hermitian symmetry, has been proposed. This also comes at the cost of 50\% spectral efficiency loss. To address this issue, a technique referred to as phase conjugated pilots has been proposed for COOFDM system \cite{STLe}. This scheme allows the spectral redundancy to be adjusted (up to 50\%) according to the required performance gain. A modification of the PCTW technique, called dual PCTW, has been proposed for single carrier systems with quadrature pulse shaping \cite{TYoshida}. This technique effectively solves the spectral redundancy problem of the PCTW technique. Recently, the idea of dual PCTW has been adapted for CO-OFDM systems by encoding the adjacent subcarrier pairs and coherently superimposing them at the receiver, to cancel the nonlinear distortions. This technique is referred to as the phase conjugated subcarrier coding (PCSC) \cite{S. T. Le}. This scheme can be effectively applied for the nonlinearity compensation without spectral efficiency loss. However, the PCSC technique does not provide any significant performance improvement without 50\% pre-EDC in the transmission link \cite{S. T. Le}. This is essentially a limiting factor for the implementation of the PCSC scheme in a dynamically routed optical network, where the exact dispersion symmetry in the link is difficult to achieve \cite{X. Liu}. 

To address the issue of the spectral redundancy associated with the PCTW technique and the requirement of dispersion
symmetry for the PCSC technique, we propose a scheme which linearly combines the data symbols on the two adjacent
subcarriers of the OFDM symbol, one at full amplitude and another at half amplitude. The phase conjugate pairs of such linearly coded signals are then transmitted on the same subcarriers of the two OFDM symbols on the two orthogonal polarizations, respectively. At the receiver, the information signals on the two orthogonal polarizations are coherently superimposed, to cancel the nonlinear distortions (including those from cross-polarization interactions). This technique is referred to as the linear polarization coded phase conjugated twin signals (LPC-PCTS). This scheme can be considered as a combination of the PCTW and PCSC techniques. The LPC-PCTS technique can be effectively used for the nonlinearity mitigation in CO-OFDM system without sacrificing the capacity of a polarization multiplexed optical transmission system. In comparison with the PCSC technique, the LPC-PCTS scheme provides a Q-factor improvement of 2.3 dB with pre-EDC and 1.7 dB without pre-EDC, respectively, at a launch power of 3 dBm for an uncompensated transmission distance of 2800 km of standard single-mode fiber (SSMF), with a net data rate of 200 Gb/s. In addition to this, the proposed LPC-PCTS technique shows a significant performance improvement when compared to the 16-QAM with PCTW technique, having the same spectral efficiency.

This paper is organized as follows. Section II discusses the concept of the PCTW technique, Section III provides the
encoding and decoding of the LPC and the nature of the nonlinear distortion cancellation through the coherent
superposition. Section IV presents the simulation results and finally, Section V reports the conclusion.          
\vspace{-0.2cm}
\section{The Concept of PCTWs} 
In PCTW technique, mutually phase conjugated twin waves are transmitted on the two orthogonal polarization states of the fiber. After transmission, the nonlinear distortions imposed on the two PCTWs, which are modulated onto a same optical carrier, are essentially anti-correlated, when a dispersion symmetry condition is met in the link \cite{XLiu}. The coherent superposition of the received PCTWs thus leads to the cancellation of the nonlinear distortions, at the expense of halving the overall spectral efficiency of a polarization multiplexed optical transmission system. Fig. 1 illustrates the basic
principle of the PCTW-based nonlinear distortion cancellation in a polarization division multiplexed (PDM) optical transmission system \cite{XLiu}. $E_{x}$, $E_{y}$ represents the transmitted electric fields and $E_{x}^{R_{x}}$, $E_{y}^{R_{y}}$ represents the received electric fields, respectively. Inset (a) and (b) shows the received signal constellations on the $x$ and $y$ polarizations, respectively, after transmission through a 2800 km SSMF. The coherent superposition of these received symbols on the two orthogonal polarizations leads to the cancellation of the nonlinear distortions and the resultant constellation is shown in the inset (c). Evidently, the constellation quality is much improved through the coherent superposition of the two PCTWs.
\begin{figure}[H]
	\centering{}\includegraphics[width=0.8\columnwidth,height=0.135\paperheight]{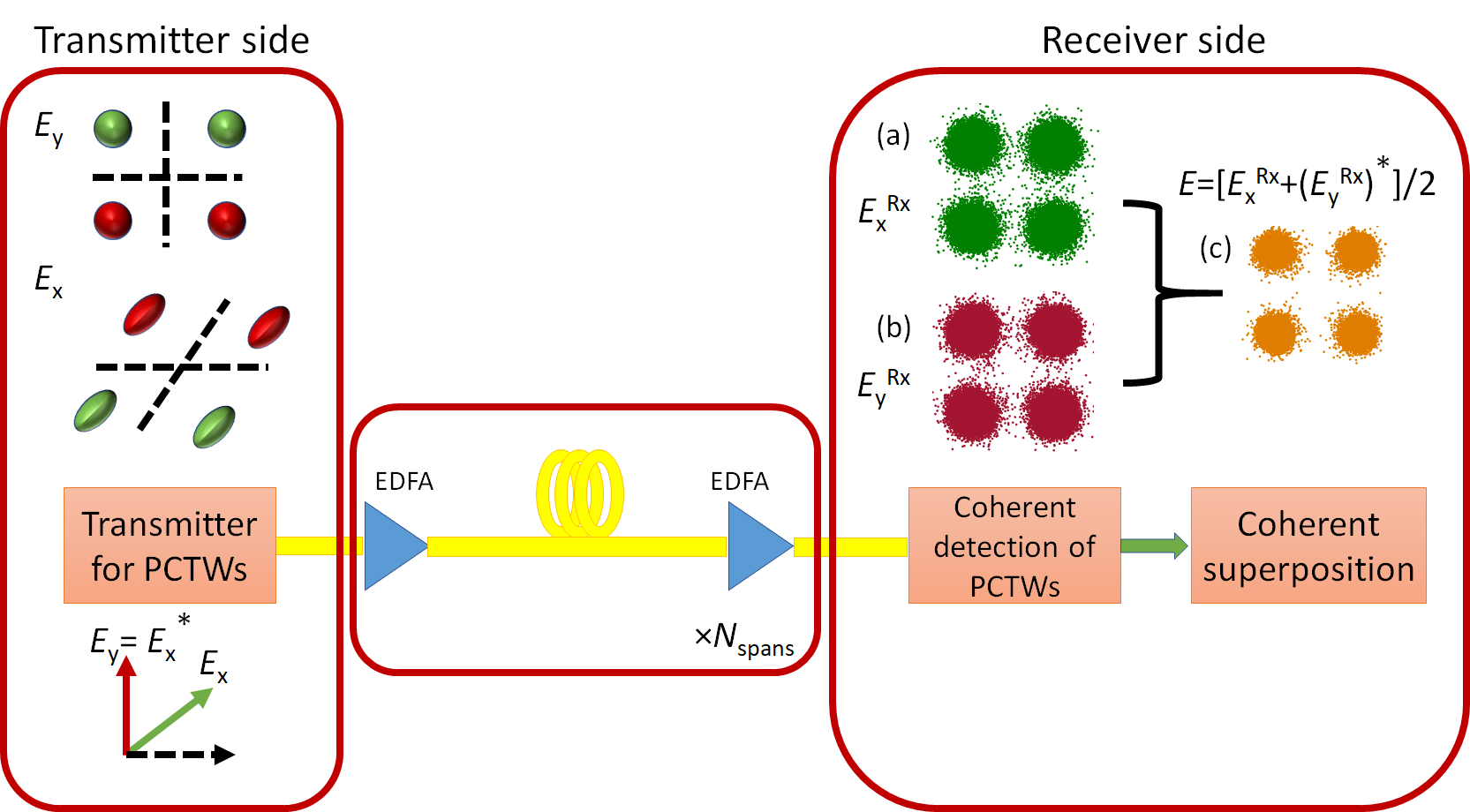}\caption{Illustration showing the cancellation of nonlinear distortions via PCTWs. EDFA: erbium doped fiber amplifier.}
\end{figure} 

It is well reported in \cite{X. Liu} that the performance gain (or sensitivity gain) obtained through the PCTW technique in the linear regime is 3 dB. This is because of halving the variance of the linear noise resulting from the amplified spontaneous emission (ASE), by coherently mixing two PCTWs. Whereas in the nonlinear regime, it can be well beyond 5 dB owing to the first-order cancellation of the nonlinear distortions through the coherent superposition of the two PCTWs \cite{X. Liu}.
\begin{multline}
	\left[\frac{\partial}{\partial z}+\frac{\left(\alpha(z)-g(z)\right)}{2}+i\frac{\beta_{2}(z)}{2}\frac{\partial^{2}}{\partial t^{2}}\right]E_{x,y}(z,t)\\
	=i\frac{8}{9}\gamma\left(\left|E_{x,y}(z,t)\right|^{2}+\left|E_{y,x}(z,t)\right|^{2}\right)E_{x,y}(z,t),
\end{multline}
where z, $\alpha$, g, $\beta_{2}$, and $\gamma$, respectively, are the propagation distance, the attenuation coefficient, the gain coefficient, the group-velocity dispersion coefficient, and the fiber Kerr nonlinearity coefficient along the transmission link. An approximate solution of (1) is obtained analytically with the application of a perturbation theory by treating $\gamma$ as a small perturbation parameter. After transmitting a distance L through the fiber, the perturbative nonlinear distortions are added to the transmitted signal fields and can be represented in the frequency domain (to first-order) as \cite{X. Liu2014}:
\begin{align}
	\delta E_{x,y}(L,w) & =i\frac{8}{9}\gamma P_{0}L_{eff}\intop_{-\infty}^{\infty}dw_{1}\intop_{-\infty}^{\infty}dw_{2}\eta(w_{1}w_{2})\nonumber \\
	& \left[E_{x,y}(w_{1})E_{x,y}(w_{2})E_{x,y}^{*}(w_{1}+w_{1}-w)\right.\nonumber \\
	& \left.+E_{y,x}(w_{1})E_{x,y}(w_{2})E_{y,x}^{*}(w_{1}+w_{1}-w)\right],
\end{align}
where $P_{0}$ is the peak optical power, $E_{x,y}(w)=\intop_{-\infty}^{\infty}E_{x,y}(0,t)e^{-iwt}dt/\sqrt{2\pi},\,L_{eff}=\left(1-e^{-\alpha L}\right)/\alpha$ is the effective length, * stands
for the phase conjugation and $\eta(w_{1}w_{2})$ is a nonlinear transfer function, defined as:
\begin{equation}
	\eta(w_{1}w_{2})=\intop_{0}^{L}\exp(G(z)-iw_{1}w_{2}C(z))dz,
\end{equation}
where $G(z)=\intop_{0}^{z}\left[g(z')-\alpha(z')\right]dz',\,C(z)=\intop_{0}^{z}\beta_{2}(z')dz'$ respectively, are the logarithmic signal power evolution and the cumulative dispersion along the optical fiber link.

When an anti-symmetric dispersion map and a symmetric power map is applied in the link such that $C(z)=-C(L-z)$ and $G(z)=G(L-z)$ then (3) becomes real-valued, i.e., $\eta(w_{1}w_{2})\thickapprox\eta(w_{1}w_{2})^{*}$. Note that the symmetric power map in the transmission link can be considered as a loose requirement owing to the low loss profile of the silica optical fibers \cite{X. Liu2014}. Since, in PCTW technique the signal and its phase conjugate are transmitted on the two orthogonal polarizations, the nonlinear distortions added onto two PCTWs are essentially anti-correlated. This eventually leads to the cancellation of the nonlinear distortions, upon coherent superposition of the received PCTWs at the receiver. In summary, by transmitting the complex conjugate of a signal on one of the orthogonal polarizations of the fiber and superimposing the received signals at the receiver results in the cancellation of the nonlinear distortions to the first-order, at the cost of halving the overall spectral efficiency of the link.

\section{The Proposed LPC-PCTS Technique} 
 
In the proposed LPC-PCTS technique, the data symbols on the adjacent subcarriers of the OFDM symbol are linearly
combined, one at full amplitude and the other at half amplitude as shown in Fig. 2. This technique is an alternate way of generating sixteen constellation points from two quadrature phase shift keying (QPSK) symbols. The constellation symbols on the right most scatter plot in Fig. 2 shows the mapping of the lower amplitude QPSK constellation symbols corresponding to a constellation symbol on the higher amplitude QPSK constellation. It is important to note that the generation of sixteen constellation symbols in this way provides an equal distance ‘d’ between the pair of constellation
points in the signal space diagram. This reduces the average probability of symbol error after detection.
\begin{figure}[H]
	\centering{}\includegraphics[width=1\columnwidth,height=0.12\paperheight]{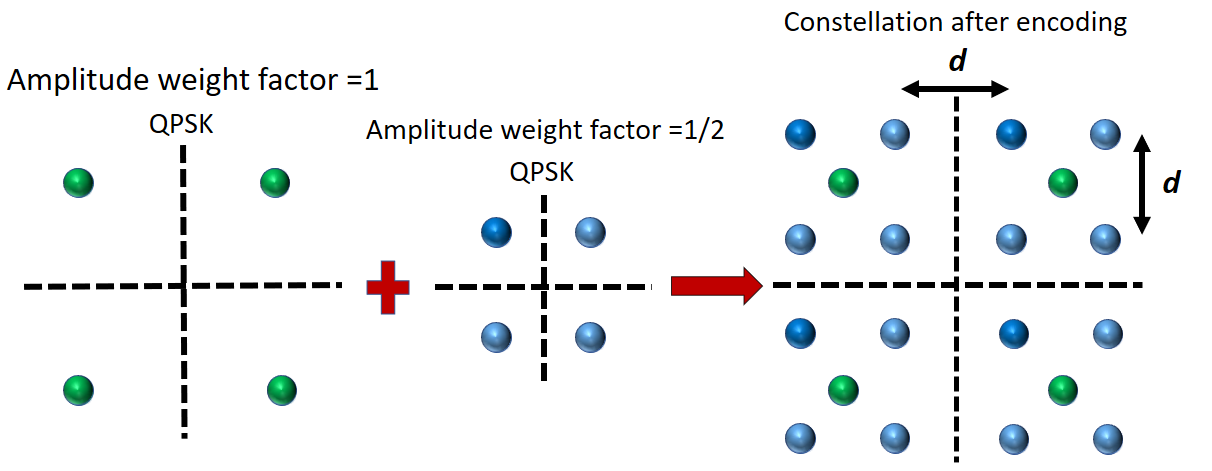}\caption{Generation of sixteen constellation symbols by combining two QPSK symbols.}
\end{figure}

At the encoder, each pair of the neighbouring OFDM subcarriers (with the indices of $2k-1$ and $2k$, where $k$ is an integer number) is encoded (see Fig. 3 (a)) as:
\begin{align}
	S_{x}(k) & =A(2k-1)+A(2k)/2\nonumber \\
	S_{y}(k) & =S_{x}^{*}(k)=A^{*}(2k-1)+A^{*}(2k)/2,
\end{align}
where $k=1,\,2,\,...,\,N/2,$ $N$ is the subcarrier number, $A$ and $S_{x/y}$ are the OFDM symbols before and after the encoding process, and the subscripts $x$ and $y$ represents the two orthogonal polarization states of the fiber.
\begin{figure}[H]
	\centering{}\includegraphics[width=1\columnwidth,height=0.08\paperheight]{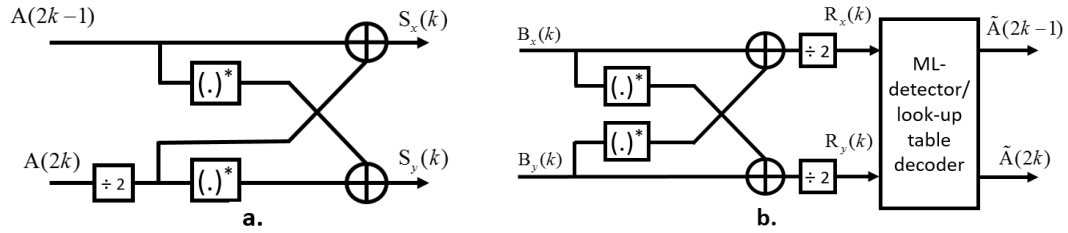}\caption{The proposed LPC-PCTS technique for CO-OFDM transmission: (a) encoder, (b) decoder.}
\end{figure}

On the other hand, if the two component QPSK symbols are combined with any other arbitrary weighted amplitude
values, then the distance between the pair of constellation points in the resultant signal space diagram is not uniform (not equal) and leads to the increased average probability of symbol error. Fig. 4 shows the received constellations for different arbitrary amplitude values for the second QPSK constellation symbols.
\begin{figure}[H]
	\centering{}\includegraphics[width=1\columnwidth,height=0.11\paperheight]{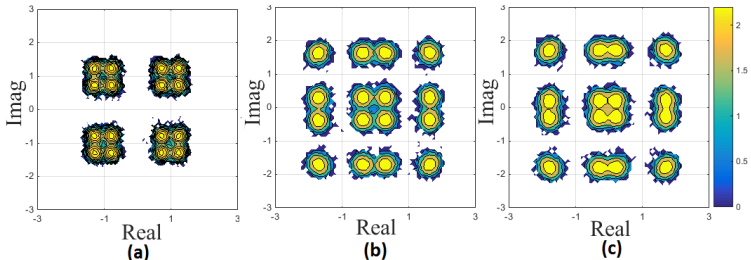}\caption{The received constellations for different amplitude values for the second QPSK symbol: (a) $1/4$, (b) $2/3$, and (c) $3/4$.}
\end{figure}

The proposed technique can be considered as a one-by-one mapping scheme in which the encoder output is essentially
a code word from a finite code alphabet. After encoding, the linearly coded symbols, $S_{x}$ and $S_{y}$ are modulated onto the electric fields corresponding to the two polarizations of the optical signal and transmitted through the fiber. The corresponding transmitted vector fields can be represented as $\left[E_{x}(0,w)\,\,\,E_{y}(0,w)\right]^{\dagger}$, where $w$ is the frequency. This feature
confirms the PCTW technique, in which the transmitted encoded signals on the two orthogonal polarizations are phase conjugate twins. This is one of the important distinctions when compared with the PCSC coding scheme, and is essential for the effective cancellation of the nonlinear distortions, upon the coherent mixing at the receiver. This will be explained in detail later in this section.

The received vector fields after propagating through the fiber can be represented as $\left[E_{x}(L,w)\,\,\,E_{y}(L,w)\right]^{\dagger}$, where $L$ is the transmission distance. After analog-to-digital conversion, the information symbols on the subcarriers corresponding to the two orthogonal polarizations are coherently superimposed as shown in Fig. 3 (b) as:
\begin{align}
	R_{x}(k) & =\left(B_{x}(k)+B_{y}^{*}(k)\right)/2\nonumber \\
	R_{y}(k) & =\left(B_{x}^{*}(k)+B_{y}(k)\right)/2,
\end{align}
where
$B_{x/y}$ and $R_{x/y}$ are the OFDM symbols before and after the decoding process, respectively.

After the coherent mixing process, the recovered symbol vector $\left[R_{x}(k)\,\,\,R_{y}(k)\right]^{\dagger}$ is passed onto the maximum-likelihood (ML)/ look-up table decoder block, as shown in Fig. 3 (b). The ML-detector calculates the distance metric on a symbolby-symbol basis and selects the code word corresponding to the minimum distance value from the set of all possible linear polarization codes. Then, a look-up table at the receiver is used for decoding the actual data symbols $\tilde{A}(2k-1)$ and $\tilde{A}(2k)$, as shown in Fig. 3 (b)

Unlike the PCSC technique, the linearly coded signals in the proposed LPC-PCTS technique are transmitted as phase
conjugate pairs on the two orthogonal polarizations. Thus, the nonlinear distortion field added onto two signal pairs are anti-correlated, when compared to PCSC technique. This leads to the full cancellation of the nonlinear distortion fields upon coherent superposition at the receiver. One disadvantage is that the modified signal constellation after the LPC coding consists of sixteen points with equal probabilities for all the points, as shown in Fig. 2. This feature drops the performance of the proposed LPC-PCTS technique in the linear (or weakly) nonlinear transmission regime, when compared to the 4-QAM with PCSC coding and the PDM-4-QAM. In the linear (or weakly) nonlinear transmission regime the performance is limited by the optical signal-to-noise (OSNR) penalty. However, the numerical simulation results show that the proposed LPC-PCTS technique outperforms the other two techniques in the highly nonlinear transmission regime, where the penalties due to the nonlinearities dominate over the OSNR penalty. This performance gain comes from the nature of the nonlinear distortion cancellation through the coherent mixing process of the proposed LPC-PCTS technique.

\subsection{The Nonlinear Distortion Cancellation}

The basic idea of the proposed LPC-PCTS technique is illustrated in Fig. 5 for the first subcarrier of the transmitted OFDM symbols on the two orthogonal polarizations, as an example.
\begin{figure}[H]
	\centering{}\includegraphics[width=1\columnwidth,height=0.19\paperheight]{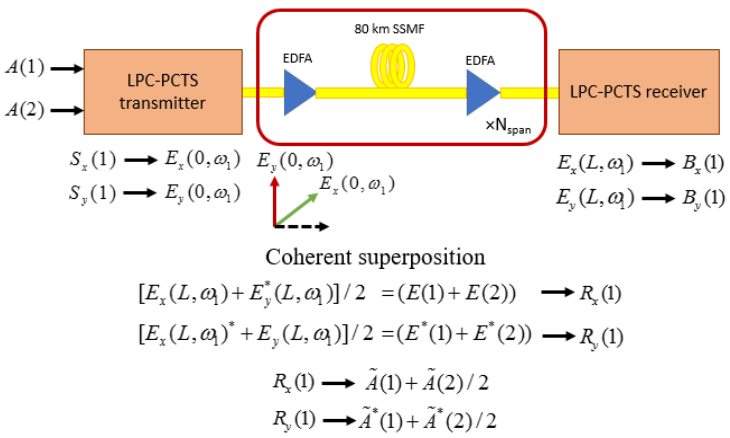}\caption{Illustration showing the LPC-PCTS technique for the first subcarrier of the transmitted OFDM symbols on the two orthogonal polarizations.}
\end{figure}

The data symbols corresponding to the first and the second subcarriers of the OFDM symbol $A$ is linearly combined
and transmitted as phase conjugate pairs on the two orthogonal polarizations. The corresponding transmitted vector fields after modulating onto the optical signal can be represented as $\left[E_{x}(0,w_{1})\,\,\,E_{y}(0,w_{1})\right]^{\dagger}$, where $w_{1}$ is the subcarrier frequency (first subcarrier). Upon transmission through the polarization division multiplexed dispersive nonlinear fiber channel, the perturbative nonlinear distortions are added to the signal tributaries of the transmitted vector fields and can be represented as:
\begin{align}
	E_{x}(L,w_{1}) & =\left(E(1)+\delta E(1)+E(2)+\delta E(2)\right)\nonumber \\
	E_{y}(L,w_{1}) & =\left(E^{*}(1)+\delta E'(1)+E^{*}(2)+\delta E'(2)\right),
\end{align}
where $E_{x/y}$ is the electric field corresponding to the data symbols on the subcarriers of the transmitted OFDM symbols on the two polarizations, $E(1)$ and $E(2)$ are the electric fields corresponding to the full amplitude and the half amplitude data symbols, respectively, on the OFDM symbol before encoding, $\delta E(1)$ and $\delta E(2)$ are the corresponding perturbative nonlinear distortion fields, $\delta E'(1)$ and $\delta E'(2)$ are the nonlinear distortion fields corresponding to the data symbols on the $y$ polarization. Here the subcarrier index $k$ in (4) is considered as one and the distance and the frequency variables of the electric fields on the right side of (6) are omitted for simplicity.

At the receiver, the coherent superposition yields:
\begin{align}
	\frac{\left[E_{x}(L,w_{1})+E_{y}^{*}(L,w_{1})\right]}{2} & =\left[\left(E(1)+\delta E(1)+E(2)+\delta E(2)\right)\right.\nonumber \\
	& +\left(E^{*}(1)+\delta E'(1)\right.\nonumber \\
	& \left.\left.+E^{*}(2)+\delta E'(2)\right)^{*}\right]/2\nonumber \\
	\frac{\left[E_{x}^{*}(L,w_{1})+E_{y}(L,w_{1})\right]}{2} & =\left[\left(E(1)+\delta E(1)+E(2)+\delta E(2)\right)^{*}\right.\nonumber \\
	& +\left(E^{*}(1)+\delta E'(1)\right.\nonumber \\
	& \left.\left.+E^{*}(2)+\delta E'(2)\right)\right]/2.
\end{align}

When there is a dispersion symmetry condition satisfied in the transmission link, then the perturbative nonlinear
distortions added to the two polarization data are essentially anti-correlated \cite{X. Liu2014}, i.e., $\left(\delta E'(1)\right)^{*}=-\delta E(1)$ and $\left(\delta E'(2)\right)^{*}=-\delta E(2)$, and it is straightforward to obtain:
\begin{align}
	\frac{\left[E_{x}(L,w_{1})+E_{y}^{*}(L,w_{1})\right]}{2} & =\left(E(1)+E(2)\right)\nonumber \\
	\frac{\left[E_{x}^{*}(L,w_{1})+E_{y}(L,w_{1})\right]}{2} & =\left(E^{*}(1)+E^{*}(2)\right).
\end{align}

It is important to note that the coherent superposition completely cancels the perturbative nonlinear distortions added to the transmitted signal fields, provided a dispersion symmetry condition is satisfied in the transmission link. This brings the performance gain for the proposed LPC-PCTS technique in the high nonlinear transmission regime, when compared to the recently proposed PCSC technique. On the other hand, if there is no dispersion symmetry in the transmission link, then the (imaginary) amplitudes of the nonlinear distortion terms are unequal \cite{X. Liu2014}, i.e, $\delta E'(1)\neq\delta E(1)$ and $\delta E'(2)\neq\delta E(2)$. However, the coherent superposition yields the nonlinear distortion terms on the two orthogonal polarization states, to be subtracted from each other. The resultant residual nonlinear distortion terms are less in magnitude and impart a considerably low performance penalty to the proposed LPC-PCTS technique. This unique feature helps the LPC-PCTS technique achieve a significant performance improvement in the absence of the dispersion symmetry in the transmission link, when compared to the PCSC technique. The recovered vector signal after the coherent superposition is input to the ML-detector for the symbol detection. The ML-detector calculates the Euclidean distance between the received vector signal and all the possible combinations of the linear polarization code, on a symbol-by-symbol basis and selects the one with minimum distance. Then, with the help of a look-up table, the data symbol which corresponds to the selected linear polarization code is determined.

\section{Simulation Results}

We conducted simulations for the proposed LPC-PCTS technique in a single channel 64 Gbaud PDM CO-OFDM
system with 4-QAM modulation format. The simulation set up is shown in Fig. 6. The input binary sequence is first
converted to parallel and then the symbol mapping is performed by using the 4-QAM modulation format. Subsequently, the proposed LPC coding is applied. Then, the linearly coded signal and its phase conjugate are mapped on to the same subcarriers of the two OFDM symbols on the two orthogonal polarization states of the fiber. There are 3300 data carrying subcarriers, and an inverse Fourier transform (IFFT) of size 4096 is performed to convert the signal into the time domain. There are four pilot subcarriers in each OFDM symbol and the remaining subcarriers are zero for oversampling. The cyclic prefix considered for the OFDM symbol is 3\%. Therefore, the net bit rate considered is approximately 200 Gb/s. Since the primary focus of this study is on the compensation of the fiber Kerr nonlinearity, the polarization mode dispersion (PMD) and polarization dependent loss (PDL) are set to zero in the simulation. The longhaul fiber link consists of 35 spans of SSMF, each having a length of 80 km, the attenuation coefficient of 0.2 dB/km, the nonlinear index coefficient of $2.4\times10^{-20} m^{2}/W$, and the dispersion coefficient of 16 ps/nm/km. The optical power loss for each span is compensated by an EDFA with 16 dB gain and 4 dB noise figure. The ASE noise is added inline to ensure that the nonlinear interaction between the signal and noise is correctly captured \cite{S. T. Le}.
\begin{figure}[H]
	\centering{}\includegraphics[width=1\columnwidth,height=0.22\paperheight]{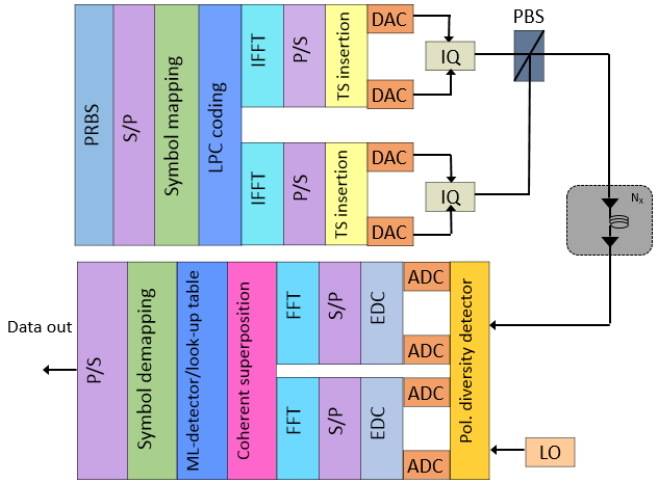}\caption{The block diagram of PDM CO-OFDM transmission system with the LPC-PCTS technique. PRBS: pseudo random binary sequence, S/P: serial/parallel conversion, P/S: parallel/serial conversion, LPC: linear polarization coding, TS: training symbol, DAC: digital-to-analog converter, ADC: analog-to-digital converter, IQ: in-phase/quadrature-phase modulator, PBS: polarization beam splitter, LO: local oscillator, EDC: electronic dispersion compensation.}
\end{figure}
At the receiver, after the polarization diversity detector, the dispersion compensation is performed using the overlapped frequency domain equalizer with the overlap-and-save algorithm \cite{C. Chen}. The channel estimation is performed with the aid of the initial training symbols (two symbols every 100 symbols). The common phase error (CPE) estimation is based on the insertion of the quasi-pilot subcarriers \cite{S. T. Le2014}. After the OFDM processing, the coherent superposition of the received information symbols on the two polarizations is performed. Then, an ML-detector/ look-up table block is used for symbol detection/decoding. Finally, the decoded symbols are demapped in the binary form. The system performance is analyzed using the $Q$-factor directly derived from the bit-error rate (BER) as \cite{S. T. Le} $Q=20\log_{10}(2\textrm{erfc}^{-1}(2\textrm{BER}))$.

Fig. 7 shows the constellations of the recovered signal for the proposed LPC-PCTS technique, after an uncompensated transmitted distance of 2800 km and an average fiber launch power of 2 dBm. Figs. 7(a) and 7(b) show the recovered constellations without pre-EDC and CPE correction, whereas Figs. 7(c) and 7(d) show the constellations with CPE correction. In Fig. 7(a), we can see a constant rotation of all the constellation points. This is due to the deterministic nonlinear phase shift
proportional to the signal intensity. This effect is absent in Fig. 7(c), where we apply the CPE correction using the quasipilot technique. It is interesting to note that the constellations (Figs. 7(b) and 7(d)) after the coherent superposition of the proposed technique, with and without the CPE correction, appear almost identical, in the absence of the dispersion symmetry in the link. There is also a significant reduction in the variance of the combined ASE and the nonlinear phase noise. This is observed through the reduced spreading of each constellation points in the signal-space diagram, after the
coherent superposition. This clearly indicates the ability of the proposed technique to compensate both the deterministic (nonlinear phase shift) and non-deterministic (nonlinear phase noise) part of the Kerr nonlinearity effects in the fiber. 
\begin{figure}[H]
	\centering{}\includegraphics[width=1\columnwidth,height=0.28\paperheight]{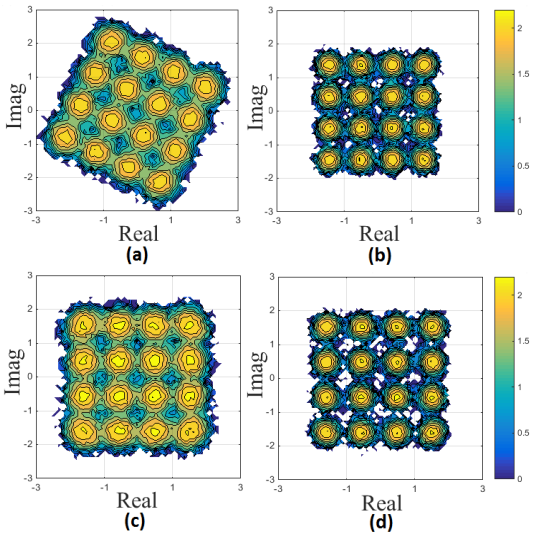}\caption{Constellations of the recovered signals for the LPC-PCTS technique without pre-EDC, after a transmission distance of 2800 km and an average launch power of 2 dBm: (a) linear comp. only (w/o CPE), (b) linear+nonlinear comp. (w/o CPE), (c) linear comp. only (w/ CPE), and (d) linear+ nonlinear comp. (w/ CPE).}
\end{figure}

Fig. 8 shows the recovered constellations, when dispersion symmetry condition is satisfied (pre-EDC) in the
transmission link. Here also we consider the simulation with and without the CPE correction, in order to emphasize the ability of the LPC-PCTS technique to compensate both the nonlinear phase shift and nonlinear phase noise, in a
symmetrically dispersion mapped optical transmission link. It is observed in Fig. 8(b) that the degree of the nonlinear distortion cancellation by the proposed LPC-PCTS technique is being enhanced through the implementation of the pre-EDC in the transmission link. This feature is due to the complete cancellation of the perturbative nonlinear distortion fields through the coherent superposition, as discussed in Section 3.1. In addition, the variance of the nonlinear phase noise in Fig. 8(d) is further decreased with the aid of the CPE correction. In summary, the proposed LPC-PCTS technique can effectively compensate for the Kerr induced nonlinear distortions, both deterministic and nondeterministic, in the presence or absence of the dispersion symmetry in the transmission link.
\begin{figure}[H]
	\centering{}\includegraphics[width=1\columnwidth,height=0.28\paperheight]{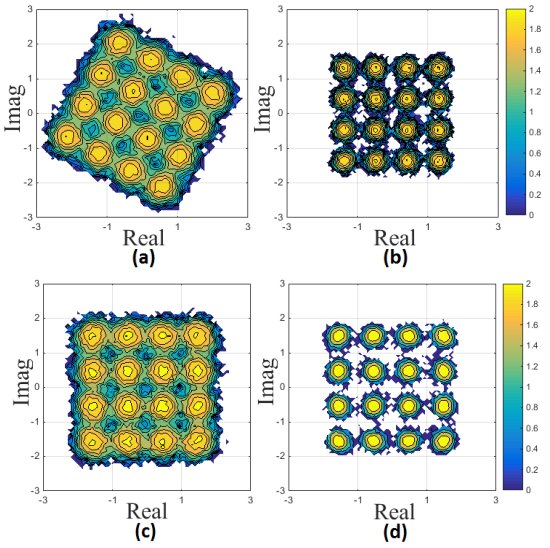}\caption{Constellations of the recovered signals for the LPC-PCTS technique with pre-EDC, after a transmission distance of 2800 km and an average launch power of 2 dBm: (a) linear comp. only (w/o CPE), (b) linear+nonlinear comp. (w/o CPE), (c) linear comp. only (w/ CPE), and (d) linear+ nonlinear comp. (w/ CPE).}
\end{figure}

The performance of the 200 Gb/s PDM-CO-OFDM system with the proposed LPC-PCTS technique and the PCSC technique are compared in Fig. 9. The cases with and without dispersion symmetry are presented to show the advantage of the proposed LPC-PCTS technique in the absence of dispersion symmetry. In this figure, the performance of PDM-4- QAM providing the same spectral efficiency is also presented. It is interesting to note that the proposed LPC-PCTS
technique outperforms the PCSC technique in the high nonlinear transmission regime, with and without the dispersion symmetry in the transmission link. It shows a $Q$-factor improvement of 2.3 dB with pre-EDC and 1.7 dB without preEDC when compared to the PCSC technique, at a launch power of 3 dBm. This performance improvement is obtained through the ability of the LPC-PCTS technique to fully cancel the nonlinear distortion fields through the coherent superposition, when there is a dispersion symmetry in the transmission link. In the absence of the dispersion symmetry, the distortion fields on the two polarizations are subtracted from each other and retains a residual distortion term, which provides comparatively low penalty to the system performance. This unique feature yields a considerable performance improvement without dispersion symmetry, when compared to the PCSC technique. On the other hand, the performance of the LPC-PCTS technique is highly limited in the linear (or weakly) nonlinear region compared to the PCSC and PDM-4-QAM schemes. This is due to the increased constellation set after the LPC coding, and thereby, the performance is limited because of the OSNR penalty. In addition, without dispersion symmetry, the PCSC technique does not provide
any significant improvement in the system’s performance and this observation confirms the results given in \cite{S. T. Le}.
\begin{figure}[H]
	\centering{}\includegraphics[width=1\columnwidth,height=0.22\paperheight]{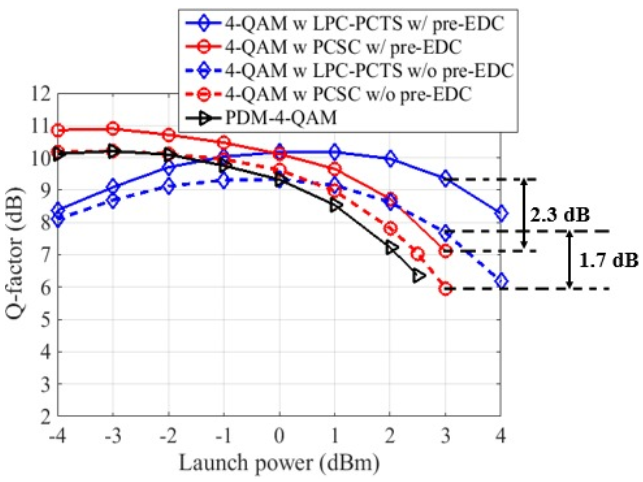}\caption{The simulation results for the LPC-PCTS and the PCSC technique for a transmission distance of 2800 km.}
\end{figure}
\begin{figure}[H]
	\centering{}\includegraphics[width=1\columnwidth,height=0.22\paperheight]{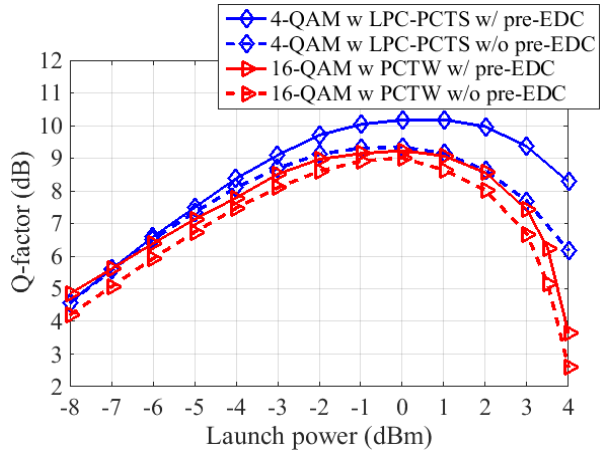}\caption{Simulation results for the LPC-PCTS and the 16-QAM w PCTW technique for a transmission distance of 2800 km.}
\end{figure}

We also conducted a simulation study for the 16-QAM combined with the PCTW technique, which provides the
same spectral efficiency as the proposed LPC-PCTS technique, and results are compared in Fig. 10. It can be seen that the proposed LPC-PCTS technique shows a significant performance improvement when compared to the 16-QAM with PCTW scheme in the presence or absence of the dispersion symmetry in the transmission link. The 16-QAM signals are highly susceptible to the fiber nonlinearities because of the increased peak-to-average power ratio after the OFDM modulation. More importantly, at high launch powers, the performance of the 16-QAM-PCTW technique is greatly limited because of the addition of the higher-order perturbative nonlinear distortion terms. The PCTW technique, in theory, is not capable of canceling the higher-order nonlinear distortion terms effectively and is experimentally demonstrated in \cite{X. Liu}.

\section{Conclusion}
We have demonstrated an efficient fiber Kerr nonlinearity compensation scheme, which combines the linear polarization coding with the phase conjugated twin signals technique, and is referred to as LPC-PCTS. In this technique, the data symbols on the adjacent subcarriers of the OFDM symbol are linearly combined, one at full amplitude and the other at a half amplitude and transmitted as phase conjugate pairs on the same subcarriers of the OFDM symbols on the two orthogonal polarizations. This technique utilizes the advantage of transmitting the phase conjugated pairs on the orthogonal polarizations, in which the nonlinear distortions added onto the signal pairs are anti-correlated. The coherent
superposition of the symbols received on the two orthogonal polarizations thus leads to the cancellation of the nonlinear distortions. The proposed technique can be effectively used for the fiber Kerr nonlinearity compensation without spectral redundancy. Numerical simulations show that the proposed LPC-PCTS technique provides a $Q$-factor improvement of 2.3 dB with pre-EDC and 1.7 dB without pre-EDC when compared to the PCSC technique, at a launch power of 3 dBm, and for an uncompensated transmission distance of 2800 km SSMF. In addition, the proposed technique also shows a significant performance improvement compared to the 16-QAM with PCTW technique, at the same spectral efficiency. In this study, we neglected the effects of PMD and PDL, which may reduce the effectiveness of the proposed LPC-PCTS technique in mitigating fiber nonlinear impairments. In future, this work can be extended to consider the impact of these imperfections on the performance of the proposed LPC-PCTS technique.


%

%


\begin{thebibliography}{1}
	
	\bibitem{REssiambre}
R. Essiambre \textit{et al}., “Capacity limits of optical fiber networks,” J. Lightw. Technol. 28(4), 662–701 (2010).

\bibitem{XLiu} 
X. Liu \textit{et al}., “Phase-conjugated twin waves for communication beyond the Kerr nonlinearity limit,” Nat. Photon (7), 560–568 (2013).

\bibitem{XYi} 
X. Yi \textit{et al}., “Digital coherent superposition of optical OFDM subcarrier pairs with Hermitian symmetry for phase noise mitigation,” Opt. Exp. (22), 13454–13459 (2014).

\bibitem{STLe}
S. T. Le \textit{et al}., “Phase-conjugated pilots for fibre nonlinearity compensation in CO-OFDM transmission”, J. Lightw. Technol. 33(7), pp. 1308–1314 (2015).

\bibitem{TYoshida}
T. Yoshida \textit{et al}., “Spectrally-efficient dual phase-conjugate twin waves with orthogonally multiplexed quadrature pulse-shaped signals,” Proc. OFC, Paper M3C.6 (2014).

\bibitem{S. T. Le}
S. T. Le \textit{et al}., “Demonstration of phase-conjugated subcarrier coding for fiber nonlinearity compensation in COOFDM transmission”, J. Lightw. Technol.33 (11), 2206–2212 (2015).

\bibitem{X. Liu}
X. Liu \textit{et al}., “Twin-wave-based optical transmission with enhanced linear and nonlinear performances”, J. Lightw. Technol.33 (5), 1037–1043 (2015).

\bibitem{X. Liu2014}
X. Liu \textit{et al}., “Fiber-nonlinearity-tolerant superchannel transmission via nonlinear noise squeezing and generalized phase-conjugated twin waves,” J. Lightw. Technol.32 (4), 766–775 (2014).

\bibitem{C. Chen}
C. Chen \textit{et al}., “Zero-guard-interval coherent optical OFDM with overlapped frequency-domain CD and PMD equalization”, Opt. Exp.19 (8), 7451–7467 (2011).

\bibitem{S. T. Le2014}
S. T. Le \textit{et al}., “Quasi-pilot aided phase noise estimation for coherent optical OFDM systems”, IEEE Photonics Technol. Lett., 26 (5), 504–507 (2014).


	
\end{thebibliography}
\end{document}